\documentclass[american, 12 pt, article]{article}

\usepackage[left=1in, right=1in, top=0.75in, bottom=0.75in]{geometry}
\usepackage[utf8]{inputenc}
\usepackage{setspace}
\usepackage{graphicx}
\usepackage{authblk}
\usepackage{sectsty}
\usepackage{gensymb}
\usepackage{parskip}
\usepackage[labelsep=period,labelfont=bf]{caption}
\usepackage{titlesec}
\usepackage{lmodern} 
\usepackage{wrapfig}
\usepackage[utf8]{inputenc}
\usepackage{chemformula}
\usepackage{siunitx}
\usepackage{xr}
\usepackage[font={small}]{caption}
\usepackage[version=4]{mhchem}
\usepackage{upgreek} 
\graphicspath{{images/}}
\sectionfont{\fontsize{13}{16}\selectfont}

\titlespacing\section{0pt}{12pt plus 2pt minus 2pt}{12pt plus 2pt minus 2pt}
\usepackage[backend=biber, style=nature] {biblatex}
\usepackage{microtype} 

\newcommand{\HTaS}{2\ch{\emph{H}–TaS2}}
\newcommand{\HNbS}{2\ch{\emph{H}–NbS2}}

\newcommand{\FexTaS}{\ch{Fe_{\emph{x}}TaS2}}

\newcommand{\FeyNbS}{\ch{Fe_{\emph{y}}NbS2}}
\newcommand{\device}{\ch{Fe_{0.42}NbS2}/\ch{Fe_{0.32}TaS2}}
\newcommand{\FexCyOz}{\ch{Fe_{\emph{x}}C_{\emph{y}}O_{\emph{z}}}}
\newcommand{\CoxCyOz}{\ch{Co_{\emph{x}}C_{\emph{y}}O_{\emph{z}}}}

\title{\Large\textbf {Tailored topotactic chemistry unlocks heterostructures of magnetic intercalation compounds}}

\author[1]{Samra Husremovi\'{c}}
\author[1]{Oscar Gonzalez}
\author[1,2]{Berit H. Goodge}
\author[1]{Lilia S. Xie}
\author[1]{Zhizhi Kong}
\author[1]{Wanlin Zhang}
\author[3]{Sae Hee Ryu}
\author[4]{Stephanie M. Ribet}
\author[4]{Karen C. Bustillo}
\author[4]{Chengyu Song}
\author[4]{Jim Ciston}
\author[5]{Takashi Taniguchi}
\author[6]{Kenji Watanabe}
\author[4]{Colin Ophus}
\author[3]{Chris Jozwiak}
\author[3]{Aaron Bostwick}
\author[3]{Eli Rotenberg}
\author[1,7,*]{D. Kwabena Bediako}

\affil[1]{\textit{Department of Chemistry, University of California, Berkeley, CA 94720, USA}}
\affil[2]{\textit{Max-Planck-Institute for Chemical Physics of Solids, N\"{o}thnitzer Str. 40, 01187, Dresden, Germany}}
\affil[3]{\textit{Advanced Light Source, Lawrence Berkeley National Laboratory, Berkeley, California 94720, United States}}
\affil[4]{\textit{National Center for Electron Microscopy, Molecular Foundry, Lawrence Berkeley National Laboratory, Berkeley, CA, USA}}
\affil[5]{\textit{Research Center for Functional Materials, National Institute for Materials Science, Tsukuba 305-0044, Japan}}
\affil[6]{\textit{International Center for Materials Nanoarchitectonics, National Institute for Materials Science, Tsukuba 305-0044, Japan}}
\affil[7]{\textit{Chemical Sciences Division, Lawrence Berkeley National Laboratory, Berkeley, CA 94720, USA}}
\affil[*]{Correspondence to: bediako@berkeley.edu}

\date{}
\begin{document}
\maketitle
\doublespacing
\textit{Abstract}

\small{\textbf{The construction of thin film heterostructures has been a widely successful archetype for fabricating materials with emergent physical properties. This strategy is of particular importance for the design of multilayer magnetic architectures in which direct interfacial spin--spin interactions between magnetic phases in dissimilar layers lead to emergent and controllable magnetic behavior. However, crystallographic incommensurability and atomic-scale interfacial disorder can severely limit the types of materials amenable to this strategy, as well as the performance of these systems. Here, we demonstrate a method for synthesizing heterostructures comprising magnetic intercalation compounds of transition metal dichalcogenides (TMDs), through directed topotactic reaction of the TMD with a metal oxide. The mechanism of the intercalation reaction enables thermally initiated intercalation of the TMD from lithographically patterned oxide films, giving access to a new family of multi-component magnetic architectures through the combination of deterministic van der Waals assembly and directed intercalation chemistry.}}
\newpage
\doublespacing

\section*{Introduction}
Solid heterostructures make it possible to leverage the proximity of disparate electronic phases to engineer exotic physical phenomena\supercite{gibert2012exchange,jiang2020concurrence,fu2020exchange}. Magnetic multilayers are among the most technologically relevant of these systems, with the pioneering discovery of giant magnetoresistance serving as a major milestone in the development of magnetic memory, sensors, and spintronic devices more broadly\supercite{baibich1988giant,heinze2011spontaneous}. Beyond elemental multilayers, the assembly of binary or ternary metal compounds into heterostructures vastly expands the range of possible functionality in such systems, as the ground state magnetic/electronic behavior of the constituent layers can be considerably more diverse\supercite{moreau2016additive,boulle2016room}. Equally important is the control over the integrity of interfaces between dissimilar materials. Interfacial atomic disorder--such as that endemic to heterostructures of some oxides and heterojunctions of polar compound semiconductors--can play a defining role in the properties of the architecture\supercite{nakagawa2006some,harrison1978polar}. 

Transition metal dichalcogenides (TMDs) intercalated with spin-bearing transition metals are appealing targets for bespoke magnetic heterostructures because they exhibit a wide range of properties that would be attractive in the design and study of magnetic multilayers: hard ferromagnetism in \ch{Fe_{0.25}TaS2}\supercite{Chen2016,morosan2007sharp,checkelsky2008anomalous,husremovic2022hard}, chiral magnetic textures in \ch{Cr_{0.33}TaS2} and \ch{Cr_{0.33}NbS2}\supercite{Miyadai1983, Togawa2012, Zhang2021,goodge2023consequences}, electrically driven collinear antiferromagnetic switching in \ch{Fe_{0.33}NbS2}\supercite{Nair2020,ic01-maniv2021exchange}, triple-Q antiferromagnetic ordering and topological Hall effect in \ch{Co_{0.33}TaS2} and \ch{Co_{0.33}NbS2}\supercite{park2023tetrahedral,takagi2023spontaneous}, and recently proposed so-called altermagnetic order in \ch{V_{0.33}NbS2}\supercite{vsmejkal2022emerging}. Their diverse properties emerge from the interplay between the host lattice structure, intercalant identity and superlattice symmetry, stoichiometry, and disorder/homogeneity\supercite{xie2022structure,goodge2023consequences}. Integrating these versatile crystals into van der Waals (vdW) heterostructures with customizable interfaces may unlock a new frontier of magnetic architectures with designed interfacial interactions and magnetoelectronic properties. However, the absence of effective synthesis methods has impeded the development of 2D heterostructures comprising magnetic intercalated TMDs. Even established techniques for fabricating heterostructures of layered crystals, such as mechanical exfoliation followed by vdW assembly, are not viable for intercalated crystals: the strong interlayer interactions imparted by the intercalants pose significant challenges in isolating thin crystals,\supercite{yamasaki2017exfoliation} which are integral components of heterostructures. Prior work on vdW assembly of intercalated TMDs has been limited to preparing magnetic tunnel junctions with incidental oxide formation at ill-defined and poorly controlled interfaces\supercite{Arai2015,yamasaki2017exfoliation}.

Here, we demonstrate a synthetic framework to create pristine heterostructures comprising magnetic TMD intercalation compounds of \HTaS\ and \HNbS\ containing Fe and Co. Previous work has shown that treatment of TMDs with solutions of zerovalent transition metals can lead to TMD intercalation compounds, with the reaction mechanism remaining elusive. We show that this topochemical reaction involves the formation of a metal oxide film derived from the zero-valent metal carbonyl precursor. We then establish that this metal oxide film releases divalent metal ions into the TMD upon annealing. Concomitantly, the oxide-exposed TMD surfaces sacrificially oxidize, thereby supplying the necessary charges to electron-dope the remaining pristine TMD layers for charge balance. Leveraging these new mechanistic insights, we develop an approach for patterning metal oxide precursor films onto TMDs that (1) facilitates studies of intercalant diffusion/ordering; (2) produces ultraclean surfaces to allow electronic structure measurements by nano-angle-resolved photoemission spectroscopy (nanoARPES); and (3) enables distinctive magnetic heterostructures of TMD intercalation compounds with atomically clean heterointerfaces. These insights and methodology establish a versatile route to new families of multi-component spintronic architectures through the combination of directed nanoscale solid-state intercalation chemistry with the myriad structural tuning knobs available for vdW heterostructures.

\begin{figure*}[htbp]
    \centerline{\includegraphics[width=\textwidth]{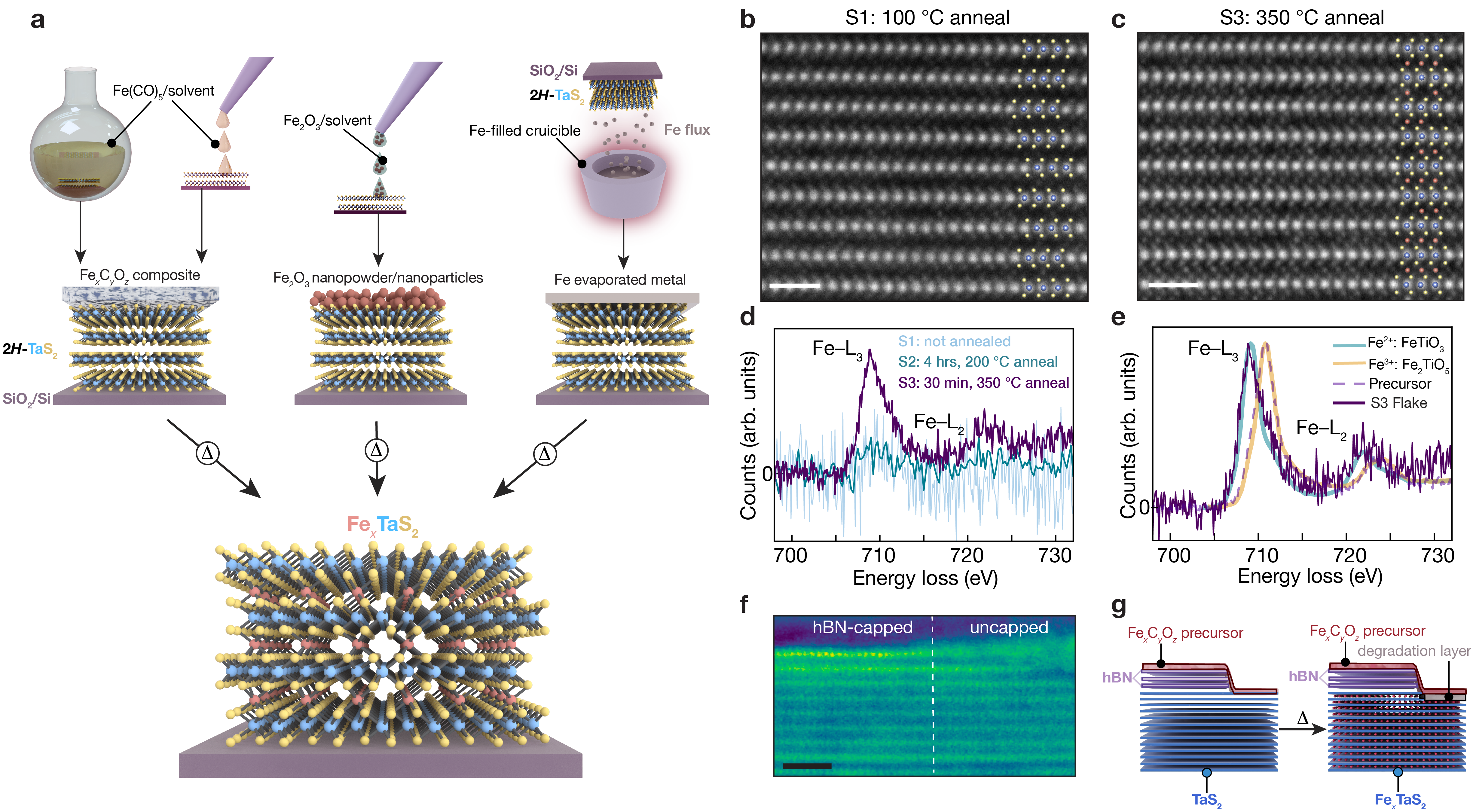}}
    \caption{\textbf{Nanoscale solid-state intercalation of vdW crystals and pristine heterostructures.} \textbf{(a)} Schematic of depositing Fe precursors onto \HTaS\ crystals on \ch{SiO2}/Si support followed by vacuum annealing to yield Fe-intercalated \HTaS\ (\FexTaS). \textbf{(b,c)} High-angle annular dark-field scanning transmission electron microscopy (HAADF-STEM) atomic-resolution images along the $[10\overline{1}0]$ zone axis of \HTaS\ flakes labeled S1 and S3 immersed in \ch{Fe(CO)5}/acetone for 24 hours at 48 \degree C and subsequently vacuum annealed at 100 \degree C for 12 hours (S1) \textbf{(b)} and 350 \degree C for 30 minutes (S3) \textbf{(c)}.  Micrographs are overlaid with the structure of \HTaS\ (ref \cite{johnston1984superconductivity}). Intercalants are marked in (c). Scale bars in (b,c): 1 nm. \textbf{(d)} Cumulative electron energy loss (EEL) spectra of \HTaS\ flakes labeled S1--S3 treated in \ch{Fe(CO)5}/acetone and vacuum annealed with distinct thermal conditions. Spectra are normalized by total acquisition time. \textbf{(e)} Cumulative EEL spectra of flake S3 and the composite \FexCyOz\ that formed during the \ch{Fe(CO)5}/acetone treatment. Experimental spectra, normalized by the L$_3$ peak maxima, are overlaid with literature spectra for \ch{(Fe^{2+})TiO3}\supercite{tan2012oxidation} and \ch{ (Fe^{3+})2TiO5}\supercite{tan2012oxidation}. Data in (d) and (e) was obtained at $\sim$100 K. \textbf{(f)} High-resolution STEM micrograph of a \HTaS\ flake that was partially capped with hexagonal boron nitride (hBN), treated with  \ch{Fe(CO)5}/isopropanol, and vacuum annealed at 350 \degree C for 1.5 hours. Scale bar: 2 nm. \textbf{(g)} Schematic of the thermally activated reaction between hBN/\HTaS\ and \FexCyOz.}
    \label{fig:Fig1}
\end{figure*}

\section*{Results}
\subsection*{\small{Synthesis of few-layer intercalation compounds with solid-state topochemistry}} 
We first illustrate the basic synthetic approach by focusing on a single intercalant species (Fe) and host lattice (\HTaS), demonstrating how nanoscale solid-state reactions offer a versatile method for synthesizing low-dimensional TMDs intercalated with transition metals. In previous work\supercite{husremovic2022hard, koski2012chemical} it has been shown that 2D flakes treated with zero-valent metal carbonyls in organic solvents (like \ch{Fe(CO)5} in acetone) (Figure 1a, left) often necessitate vacuum annealing to evince clear signs of intercalation in their Raman spectra, such as the appearance of new phonon modes associated with the intercalant superlattice and Raman peak shifts indicative of charge transfer to the layered host lattice. Indeed, we find this thermal treatment to be consistently required in the case of \HTaS\ \supercite{husremovic2022hard}. Here, to first understand the effect of this thermal treatment at an atomic level, we use atomic-resolution high-angle annular dark-field scanning transmission electron microscopy (HAADF-STEM) to examine samples heated to different temperatures. HAADF-STEM imaging was performed on cross-sectional samples made from \HTaS\ flakes on \ch{SiO2}/Si treated with 10 mM \ch{Fe(CO)5}/acetone and subjected to thermal annealing at 100 \degree C (Figure 1b), 200 \degree C or 350 \degree C (Figure 1c). These samples, labeled S1--S3, display substantial structural differences: whereas interstitial sites in the vdW interface of S1 and S2 appear largely vacant, S3 displays a high occupancy of intercalants in the pseudo-octahedral sites. To identify these intercalants and compositionally characterize S1--S3, we complement this atomic-resolution imaging with spatially resolved electron energy loss spectroscopy (EELS). EEL spectra of S1--S3 reveal the spectroscopic fingerprint of Fe only for the sample annealed at 350 \degree C (S3) (Figure 1d). Taken together, HAADF-STEM, EELS, and Raman data show that substantial intercalation does not occur during treatment in \ch{Fe(CO)5} solution alone, even with prolonged (24.5 h) soaking in a 10 mM solution of the zerovalent metal carbonyl. Instead, we observe the formation of a thin film on the surface of the chip containing Fe, C, and O. Upon annealing at 350 \degree C (in a process that necessitates exposing the film for a few minutes to ambient conditions) under high vacuum, \HTaS\ flakes in contact with this film are topochemically converted to \FexTaS\ in a solid-state reaction. The chemical details of this topochemical reaction are now discussed.

\subsection*{\small{Nature of the intercalation reaction with carbonyl-derived films}}
To probe the redox process accompanying intercalation, we examine the Fe species in the carbonyl-derived films and \HTaS\ using a combination of EELS and X-ray photoelectron spectroscopy (XPS). While Fe$^{3+}$ is the majority species in the films (Figure 1e), Fe centers intercalated between \HTaS\ layers exist exclusively as Fe$^{2+}$ (Figure 1e)\supercite{leapman1982study,tan2012oxidation}, in agreement with prior work\supercite{husremovic2022hard,eibschutz1975ferromagnetism}. We gained additional insight into the intercalation redox chemistry through solid-state reactions between \ch{TaS2} and \ch{Fe2O3} powders. These reactions reveal that Ta$^{4+}$ in \ch{TaS2} disproportionates to \ch{Ta2O5} and reduced (electron-doped) \ch{TaS2} intercalated with Fe$^{2+}$ (\textit{i.e.}, \FexTaS). A possible reaction that describes this process is:

\begin{equation}
(6\textit{x} + 10) \ce{TaS2} + 5x\ce{Fe2O3 ->} 3x \ce{Ta2O5} + \ce{10 Fe_{x}TaS2} + 12x\ce{S}
\label{eq:equation1}
\end{equation}

To test this further, we examine the reactions of carbonyl-derived metal oxide films on nano-thick \HTaS\ flakes using high-resolution STEM and EELS measurements (Figure 1f). We find that  \HTaS\ layers in direct contact with the films undergo oxidation, while crystalline \FexTaS\ layers form underneath, congruent with the proposed disproportionation. These data also show that Fe intercalation can take place through the basal plane of \HTaS, with Fe diffusing both vertically and laterally through \HTaS\ crystals. Moreover, we find that the diffusivity of Fe enables intercalants to move from regions contacting the film to areas partially encapsulated with a chemically inert hexagonal boron nitride (hBN) (Figure 1f). Notably, hBN-encapsulated regions do not contain \ce{Ta2O5} layers, demonstrating that pristine \FexTaS\ crystals can be synthesized by combining vdW heterostructuring (capping with hBN) and this nanoscale solid-state disproportionation chemistry.

To confirm and explore the versatility of this chemistry, we also vacuum annealed mechanically exfoliated \HTaS\ crystals that were coated with a suspension of \ch{Fe2O3} nanopowder/nanoparticles or a thin layer of electron-beam evaporated Fe metal that subsequently oxidizes upon air exposure according to XPS (Figure 1a, middle and right). Both routes also produce \FexTaS. In addition, we find that the carbonyl-derived metal oxide films can simply be prepared by drop-casting or spin-coating solutions of \ch{Fe(CO)5} in isopropanol and then exposing these surfaces to air (Figure 1a, left).

\subsection*{\small{Intercalation and interlamellar transport from patterned oxide films}} 
The observation that intercalation proceeds via a solid-state topotactic reaction and not a heterogeneous reaction between solid \HTaS\ and \ch{Fe(CO)5} in solution enables the precise patterning of carbonyl-derived oxide films onto 2D flakes. This precursor patterning now allows us to investigate in detail the effect of vacuum annealing conditions on the intercalation, lateral transport, and crystallographic ordering of Fe centers in few-layer \HTaS\ flakes. Nanoscale structural ordering of Fe centers was examined using temperature-dependent four-dimensional scanning transmission electron microscopy (4D-STEM). In this experiment, a $\sim$5.2 nm converged electron probe was scanned across a 1.3 $\mu$m $\times$ 0.3 $\mu$m window on an 11 nm thick \HTaS\ crystal (through a hole of diameter 2 $\mu$m in the underlying silicon nitride TEM membrane), immediately adjacent to the patterned oxide film. Electron diffraction patterns were obtained at each probe position (Figure 2a). This experiment was repeated as the temperature of the TEM holder was increased from 14 \degree C to 350 \degree C. We find that nanoscale Fe ordering commences at 250 \degree C, as evidenced by the emergence of faint $\sqrt{3} \times \sqrt{3}$ Fe superlattice diffraction peaks (Figure 2b). This is consistent with our high-resolution scanning transmission electron microscopy (HRSTEM) and electron energy loss spectroscopy (EELS) experiments: Fe was absent in samples annealed up to 200 \degree C (Figure 1b--d). 

\begin{figure*}[htbp]
    \centerline{\includegraphics[width=\textwidth]{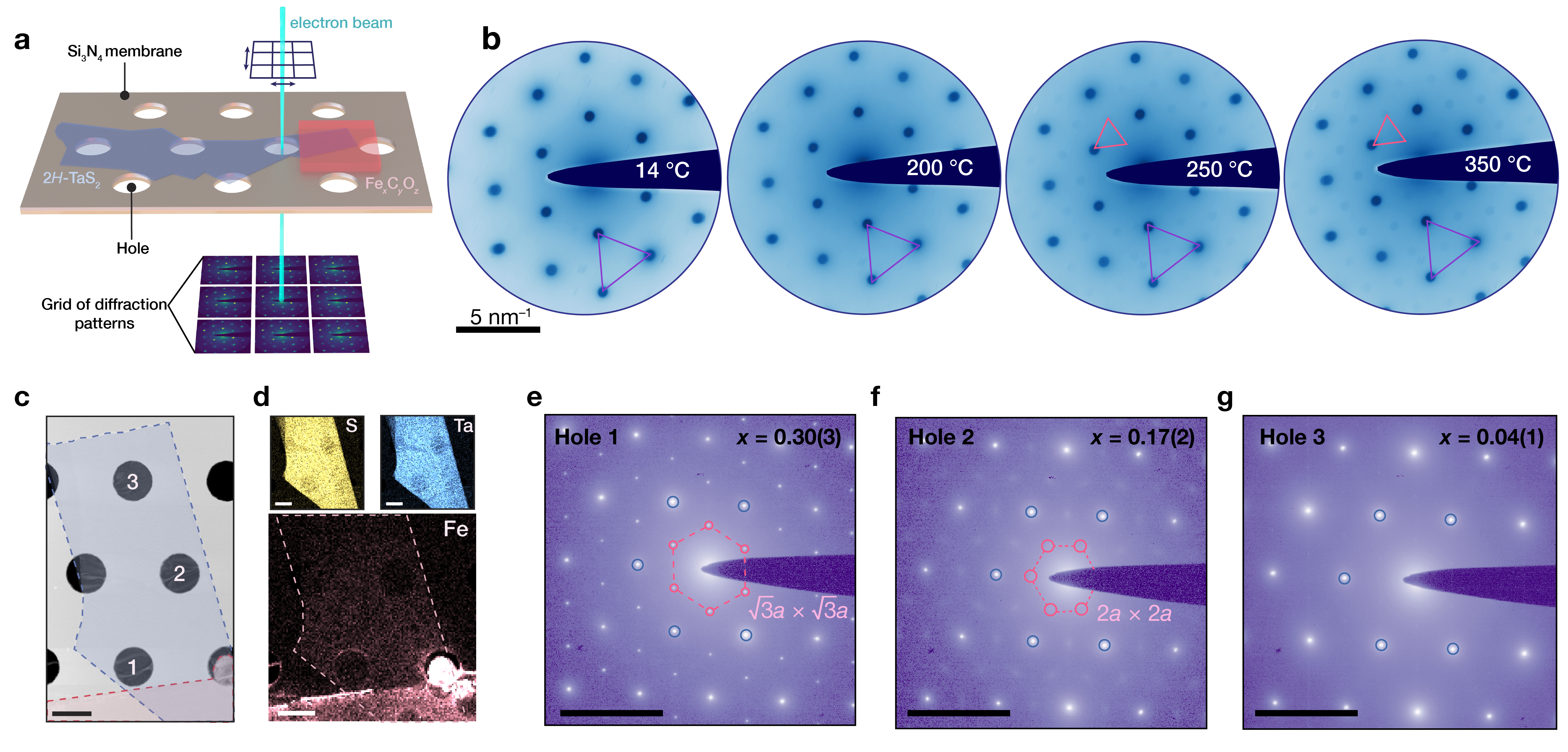}}
    \caption{\textbf{Probing uptake and diffusion of Fe in \HTaS.} \textbf{(a)} Schematic illustrating four-dimensional scanning transmission electron microscopy (4D-STEM) of a \HTaS\ flake with a patterned \FexCyOz\ precursor. \textbf{(b)} Mean diffraction patterns, displayed on a logarithmic scale, of 4D-STEM datasets obtained during in-situ heating of a \HTaS\ flake with a patterned \FexCyOz\ precursor. Diffraction reflections of the \HTaS\ lattice and the Fe superlattice are indicated in violet and red, respectively. The precursor was made by drop-casting a solution of \ch{Fe(CO)5}/isopropanol. Data was collected along the \textit{c}-axis of the flake after a 30-minute temperature equilibration period. \textbf{(c)} Plan-view STEM micrograph ($\parallel$ \textit{c}-axis) of a \HTaS\ crystal with a patterned \FexCyOz\ precursor after ex-situ vacuum annealing for 1 hour at 350 \degree C. The \FexCyOz\ precursor was made by drop-casting a solution of \ch{Fe(CO)5}/isopropanol. The \HTaS\ crystal and the \FexCyOz\ patterned precursor are false-colored in blue and red, respectively. Number labels for the TEM grid holes are marked in white. Scale bar: 2 $\mu$m \textbf{(d)} STEM-EDS maps of sample (c). In the Fe EDS map, \HTaS\ is outlined in a dashed line. Scale bars: 2 $\mu$m. \textbf{(e--g)} Selected area electron diffraction (SAED) patterns of (c) obtained for grid hole 1 \textbf{(e)}, hole 2 \textbf{(f)}, hole 3 \textbf{(g)}. The Fe/Ta ratio (\textit{x}) is marked on the SAED patterns. First-order diffraction peaks from the \HTaS\ lattice and the Fe superlattice are highlighted in blue and red, respectively. Scale bars (e--g): 5 nm$^{-1}$.
    }
    \label{fig:Fig2}
\end{figure*}

Further, 4D-STEM analysis revealed that, as expected for any solid-state reaction with a unidirectional reactant source, the amount of ordered Fe increased with higher temperatures and longer annealing times. As these diffraction-based experiments only provide insight into the structurally ordered Fe, 4D-STEM measurements were complemented with STEM-EDS examination of a 22 nm flake annealed ex-situ (Figure 2c). Compositional analysis of this crystal at three holes in the underlying membrane (centered approximately 1, 5, and 10 $\mu$m away from the oxide film) shows a steady decrease in Fe content ($x$ of 0.30(3), 0.17(2), and 0.04(1), respectively) with increasing distance from the patterned film (Figure 2d). Additionally, selected area electron diffraction (SAED) patterns indicate a corresponding change in the Fe superlattice ordering from a well-defined $\sqrt{3} \times \sqrt{3}$ (Figure 2e) to a diffuse $2 \times 2$ superlattice (Figure 2f), ultimately disappearing entirely in hole 3 (Figure 2g). Thus, annealing \HTaS\ samples in contact with a lithographically defined Fe oxide film results in a gradient in Fe content over several microns in distance. Taken together, our findings reveal that the primary determinants for achieving the desired intercalation products are the annealing temperature, annealing time, positioning of the intercalant oxide film, and the dimensions of \HTaS\ flakes.

\subsection*{\small{Evolution of electronic band structure of thin \FexTaS\ with Fe content}} 
The electronic ramifications of Fe\textsuperscript{2+} intercalation from these oxide films remain unanswered. If our proposed disproportionation chemistry (equation 1) is correct, we would expect electron transfer to the \HTaS\ host. We turn to angle-resolved photoemission spectroscopy (ARPES) measurements with submicron beam diameter to examine the effects of Fe intercalation from \FexCyOz\ films on the electronic structure of low-dimensional \HTaS. A \FexCyOz\ film was patterned onto one end of an 11 nm \HTaS\ crystal, which was also partially encapsulated with a monolayer of hBN to prevent degradation of the measured region. This crystal was then annealed in high vacuum for 1 hour at 350 \degree C. As depicted in Figure 3a, photoemission spectra as a function of crystal momentum were acquired to measure the electronic band structure of the material over a range of distances away from the patterned \FexCyOz\ film with spatial resolution on the order of 1 $\mu$m. We also acquired core-level spectra, confirming the presence of Fe in the measured area. Representative Fermi surface and energy versus momentum cuts along the $\Upgamma$--K direction are presented in Figure 3b and Figure 3c, respectively. In measurements of bulk intercalated TMDs, cleaved surfaces can possess two terminations: intercalant termination or TMD termination \supercite{xie2023comparative,ko2011rkky,edwards2023giant}. The measured electronic features here are qualitatively consistent with a TMD-terminated surface, which provides evidence for minimal intercalation between the hBN and \textit{H}-\ch{TaS2} layers.  Interestingly, we also observe that the Fermi surface cuts obtained at different points away from the precursor exhibit distinct intensity distributions, indicating potential variations in electronic reconstruction across the crystal\supercite{xie2023comparative}.

\begin{figure*}[htbp]
    \centerline{\includegraphics[width=\textwidth]{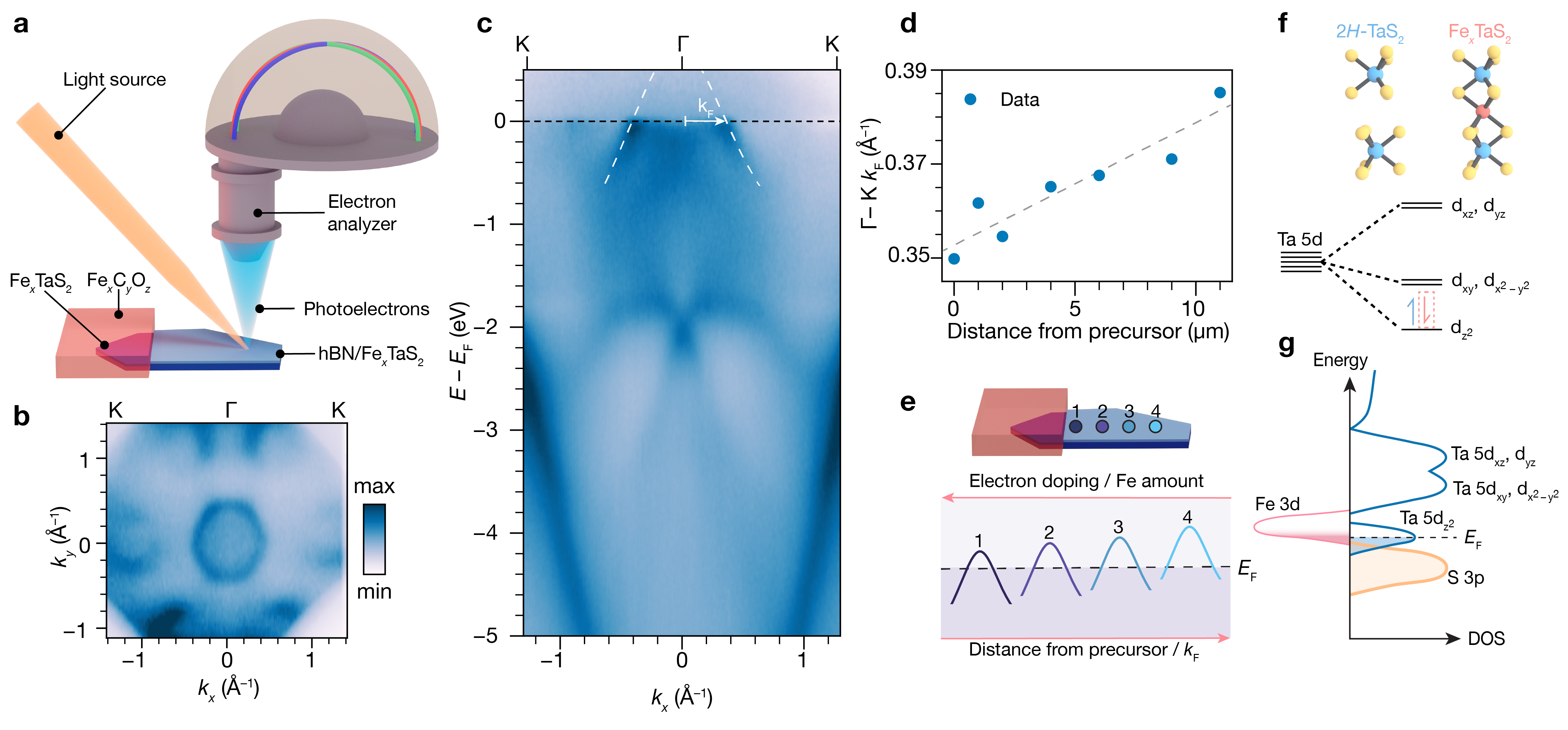}}
    \caption{\textbf{Spatially mapping the band structure of ultra-clean intercalated heterostructures with nanoARPES.} \textbf{(a)} A schematic depicting the nanoARPES experiment conducted on a \FexTaS\ encapsulated with monolayer hBN. The sample was prepared by selectively applying \FexCyOz\ onto the area of \HTaS\ not covered by hBN, followed by vacuum annealing. \textbf{(b,c)} Normalized ARPES Fermi surface \textbf{(b)} and ARPES band dispersion along the $\Upgamma$--K direction \textbf{(c)} of a hBN/\FexTaS\ heterostructure at a 4 $\mu$m distance from the patterned \FexCyOz\ precursor. In (c), the Fermi wavevector (\textit{k}\textsubscript{F}) is marked, denoting where the band forming the hole pocket around $\Upgamma$ intersects the Fermi level (\textit{E}\textsubscript{F}). This band is marked by a white dashed line. Data in (b-c) was obtained at 19 K with h$\nu$ = 118 eV and linear horizontal (LH) polarization. \textbf{(d)} \textit{k}\textsubscript{F} along the $\Upgamma$--K direction extracted from ARPES band-dispersions obtained at different distances from the \FexCyOz\ precursor. The \textit{k}\textsubscript{F}  values were obtained by fitting the momentum distribution curves (MDCs) at the Fermi level (\textit{E}\textsubscript{F}) to Lorentzians. The dashed gray line serves as a visual guide. \textbf{(e)} Schematic of the band forming the hole pocket at $\Upgamma$ as the distance from the precursor increases from spot 1 to 4. \textbf{(f)} Qualitative d-orbital splitting diagram for the trigonal prismatic Ta center in \FexTaS. A dashed electron in the d\textsubscript{z\textsuperscript{2}} orbital denotes additional electron filling upon Fe intercalation, concomitant with charge transfer to \HTaS. \textbf{(g)} Qualitative representation of the density of states of Fe-intercalated \HTaS.}
    \label{fig:Fig3}
\end{figure*}

A quantitative analysis of the nanoARPES datasets collected at various spots across the sample shows the evolution of the band structure with distance from the oxide film. By fitting the momentum distribution curves (MDCs) at the Fermi level (\textit{E}\textsubscript{F}) along $\Upgamma$--K to Lorentzians, we extracted the Fermi wavevector (\textit{k}\textsubscript{F}) to quantify the size of the hole pocket around $\Upgamma$. This analysis reveals that the diameter of the central hole pocket increases as a function of distance away from the oxide film, with \textit{k}\textsubscript{F} increasing from 0.35 \r{A}$^{-1}$ to 0.39 \r{A}$^{-1}$ (Figure 3d). This increase in \textit{k}\textsubscript{F},  which is inversely proportional to electron filling, reveals a decrease in electron doping of the \ch{TaS2} with distance away from the oxide film (Figure 3e). This doping trend mirrors the Fe concentration gradient uncovered by Raman mapping of this sample, as well as the STEM EDS and electron diffraction measurements of the sample presented in Figure 2c--g. As distance from oxide film correlates with the amount of intercalated Fe, these data provide a highly informative picture of the progress of the intercalation reaction itself. As Fe\textsuperscript{2+} is intercalated, the \HTaS\ host lattice is progressively electron-doped. In a simplified picture, this electron transfer to \HTaS\ results in an increased population of the band consisting largely of Ta \textit{d}\textsubscript{z\textsuperscript{2}} orbital character, which is at the Fermi level (Figure 3f,g)\supercite{xie2022structure}.  Although the experimental band structures of TMDs involve more complexity than this basic explanation\supercite{xie2023comparative,mattheiss1973band}, increasing Fe\textsuperscript{2+} content leads to increased electron doping that in turn reduces the size of the central hole pocket of this Ta \textit{d}\textsubscript{z\textsuperscript{2}} band (Figure 3e). Therefore, the data of Figures 2 and 3 show how the Fe\textsuperscript{2+} content has structural and electronic implications, and the interplay between these factors is crucial for explaining the emergent magnetism in these materials\supercite{xie2022structure}.

\begin{figure*}[htbp]
    \centerline{\includegraphics[width=\textwidth]{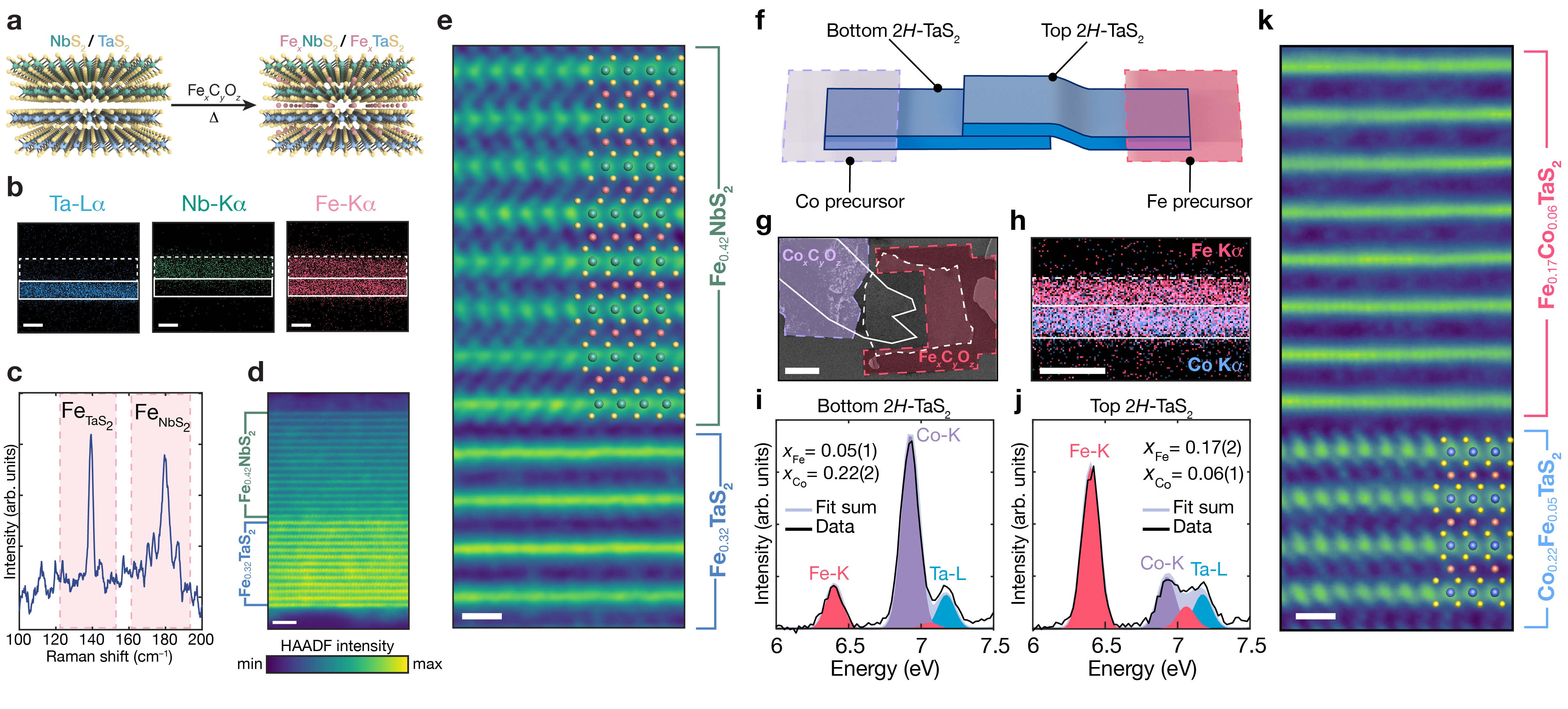}}
    \caption{\textbf{Synthesis and characterization of intercalated heterostructures.}  \textbf{(a)} Schematic illustrating the synthesis of Fe-intercalated heterostructure of \HNbS/\HTaS. \textbf{(b)} STEM energy-dispersive X-ray spectroscopy (EDS) maps perpendicular to the \textit{c}-axis of the heterostructure synthesized as depicted in (a). The dashed and solid white lines mark the edges of \HNbS\ and \HTaS, respectively. Scale bars: 10 nm. \textbf{(c)} Ultra-low frequency (ULF) Raman spectra of the heterostructure from (b). \textbf{(d)} High-resolution STEM image of the Fe$_{0.42}$NbS$_2$/Fe$_{0.32}$TaS$_2$ heterostructure from (b) along the $[10\overline{1}0]$ zone axis of the \ch{TaS2} flake. The azimuthal misalignment (twist) angle is 3.5\degree, determined using Kikuchi bands. Scale bar: 2 nm. \textbf{(e)} Atomic-resolution HAADF-STEM image of the heterostructure from (b) obtained along the $[10\overline{1}0]$ zone axis of the \ch{NbS2} flake. Scale bar: 5 \r{A}. Crystal structure of \ch{Fe1\textsubscript{/}3NbS2} from ref \cite{Anzenhofer1970} is overlaid with the \HNbS\ flake. \textbf{(f)} Illustration of a heterostructure comprising two \HTaS\ flakes with the top and bottom one partially covered with the \FexCyOz\ and \CoxCyOz\ precursors, respectively. \textbf{(g)} Plan-view ($\parallel$ \textit{c}-axis) scanning electron microscope (SEM) image of a heterostructure made according to the synthetic design in (f). The heterostructure with the patterned precursors was annealed for 1 hour at 350 \degree C. In the SEM image, \HTaS\ flakes are outlined in white, while the Co and Fe precursors are false-colored in violet and red, respectively. Scale bar: 5 $\mu$m. \textbf{(h)} STEM-EDS map of a cross-sectional TEM sample made from heterostructure (g). Scale bar: 20 nm. \textbf{(i,j)} Cumulative STEM-EDS spectra of the bottom \textbf{(i)} and top \textbf{(j)} \HTaS\ flakes from (g). In (i,j), marked \textit{x}\textsubscript{\textit{M}}, where \textit{M} = (Fe, Co), denotes the stoichiometric ratio between \textit{M} and Ta. \textbf{(k)} Atomic-resolution HAADF-STEM image of the heterostructure from (g) obtained along the $[10\overline{1}0]$ zone axis of the bottom \HTaS\ flake. The azimuthal angle (twist) between the \HTaS\ flakes is 20\degree. Crystal structure of \ch{Fe1\textsubscript{/}3TaS2} from ref \cite{wijngaard1991optical} is overlaid with the bottom \ch{TaS2} flake. Scale bar: 5 \r{A}. 
    }
    \label{fig:Fig4}
\end{figure*}

\subsection*{\small{Intercalation compound heterostructures}}
Having unveiled this topotactic chemistry for intercalating TMDs with open-shell metal ions, we exploit this method to craft heterostructures of magnetic intercalation compounds. We demonstrate two paradigms for this combination of topotactic chemistry and vdW heterostructure assembly. In the first case, we intercalate a TMD heterostructure (\HTaS/\HNbS) with a single intercalant element (Fe), and in the second case, we create heterostructures using two different intercalants (Fe and Co) in a stack of the same TMD, \HTaS. 

\FexTaS\ is ferromagnetic (FM) in the bulk\supercite{morosan2007sharp,Chen2016,checkelsky2008anomalous} as well as in the few-layer limit\supercite{husremovic2022hard}. In contrast, \FeyNbS\ is known to be antiferromagnetic (AFM) in the bulk\supercite{Nair2020,ic01-maniv2021exchange,Nair2020}. Consequently, the synthesis of \FexTaS/\FeyNbS\ heterostructures with clean interfaces would enable the construction of FM/AFM bilayers that may display technologically relevant magneto-electronic properties as a result of interfacial coupling between the dissimilar magnetic states\supercite{albarakati2022electric,lachman2020exchange,kong2023near}. To synthesize clean \FexTaS/\FeyNbS, interfaces, we prepare \HTaS/\HNbS\ heterostructures using vdW assembly, deposit a \FexCyOz\ film from \ch{Fe(CO)5}/toluene solution, and anneal the heterostructure at 350 \degree C for 1.5 hours (Figure 4a). STEM-EDS mapping confirms the presence of Fe across the resulting heterostructure (Figure 4b) and estimates the Fe content to be 0.42 and 0.32 per \HNbS\ and \HTaS, respectively. The presence of low frequency Raman modes at 139 cm\textsuperscript{-1} and 181 cm\textsuperscript{-1} (Figure 4c, Extended Data Figure 1) are consistent with $\sqrt{3} \times \sqrt{3}$ Fe superlattices in \FexTaS\ and \FeyNbS, respectively \supercite{Fan2021,husremovic2022hard,nagao2001raman}. HRSTEM imaging of this Fe$_{0.32}$NbS$_2$/Fe$_{0.30}$TaS$_2$ assembly reveals an atomically flat and sharp heterointerface between the dissimilar 2D crystals (Figure 4d,e) with no amorphous oxide tunneling barrier (typically observed in heterostructures made from cleaved bulk crystals)\supercite{Arai2015,yamasaki2017exfoliation}. Moreover, atomic-resolution HRSTEM confirms that Fe ions decorate the interstitial between all vdW layers, including the 3.5-degree twisted heterointerface of Fe-intercalated \HTaS\ and \HNbS\ crystals (Figure 4e).

To create co-intercalated heterostructures, we pattern carbonyl-derived oxide films onto TMD stacks. This multi-precursor technique was applied to two stacked \HTaS\ flakes, with oxide precursors \FexCyOz\ and \CoxCyOz\ films patterned on the top and bottom \HTaS\ crystals, respectively (Figure 4f,g). After thermal annealing at 350 \degree C, we examine the overlapping region of the \HTaS\ flakes using STEM-EDS (Figure 4h--j). These measurements show that the top flake is principally Fe-rich and the bottom flake is Co-rich. Notably, although Fe and Co are present in both flakes, each flake contained a substantially higher concentration of the metal patterned directly onto it, implying that the lateral diffusion is considerably more facile than the vertical diffusion. This observed intercalant segregation distinguishes these heterostructures from bulk crystals grown with solid-state methods, where multiple metal intercalants disperse uniformly throughout the host lattice\supercite{pan2023fe}. Atomic-resolution HAADF-STEM imaging of our heterostructures provides an explanation for this segregation (Figure 4k), showing that the interface between the \HTaS\ crystals is atomically sharp and, importantly, azimuthally misaligned (twisted). Twisting creates an in-plane moir\'e superlattice\supercite{VanWinkle2023,weston2022interfacial} and necessarily reduces the availability of optimal pseudo-octahedral intercalant sites. These results raise the possibility that controlling these azimuthal angles may be used to shape the gradient of co-intercalants and the resulting magnetism in tailored mixed heterostructures.

\begin{figure*}[htbp]
    \centerline{\includegraphics[width=3.5 in]{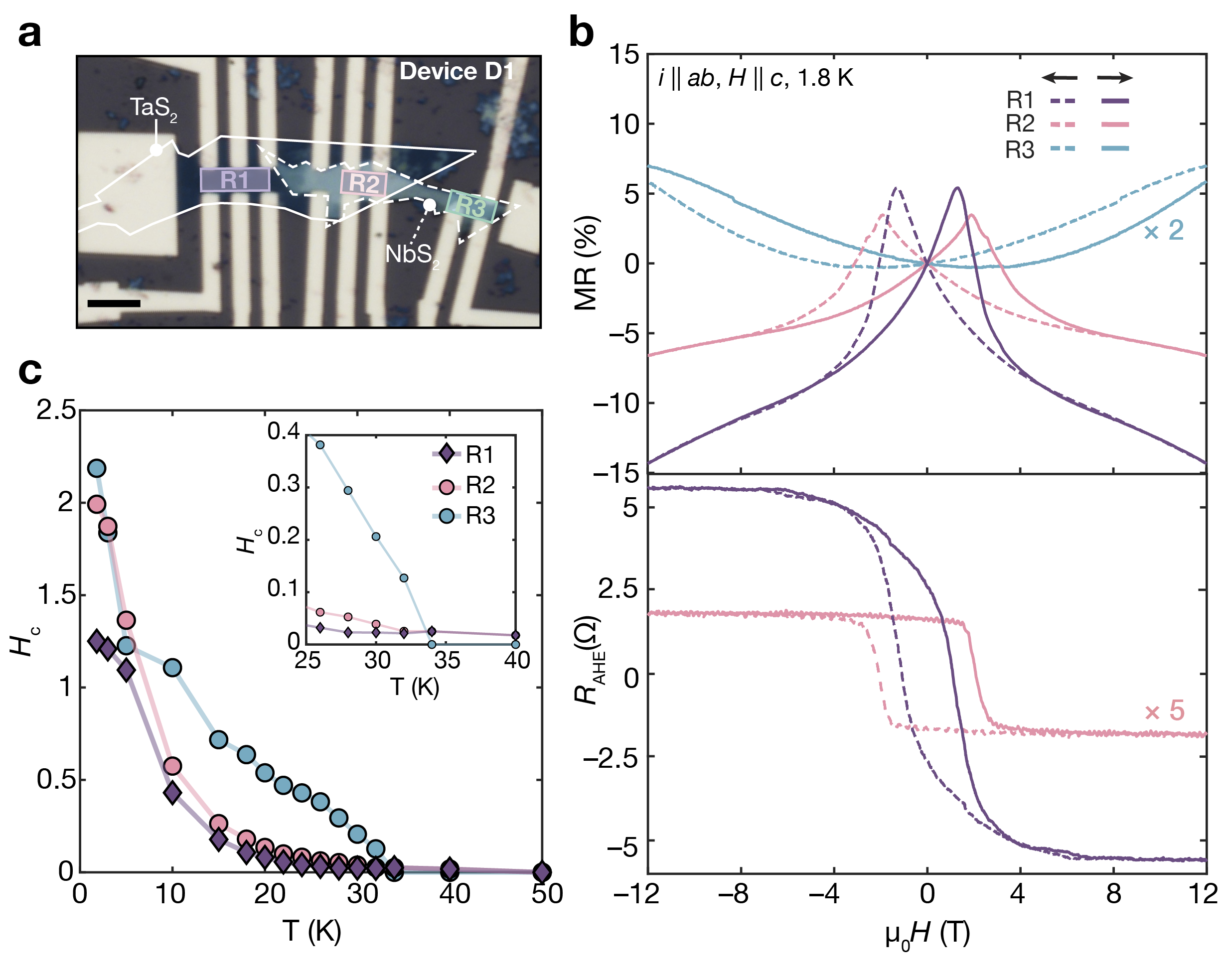}}
    \caption{\textbf{Magnetotransport of iron-intercalated \HNbS/\HTaS\ heterostructures.} \textbf{(a)} Optical micrograph of the measured mesoscopic device. Three distinct measured regions are false-colored and labeled: R1--Fe\textsubscript{0.32(1)}TaS\textsubscript{2}, R2--Fe\textsubscript{0.42(4)}NbS\textsubscript{2} / Fe\textsubscript{0.31(2)}TaS\textsubscript{2} and R3--Fe\textsubscript{0.32(3)}NbS\textsubscript{2}. \textbf{(b)} Field-dependent magnetoresistance (MR) and anomalous Hall resistance ($R_{\mathrm{AHE}}$) recorded in regions R1--R3. \textbf{(c)} Temperature-dependent coercive field ($H_c$) for R1--R3. For R1--R2, the average of $H_c$ is obtained from both MR and $R_{\mathrm{AHE}}$ measurements. Here, $H_c$ refers to the field where MR reaches its maximum while $R_{\mathrm{AHE}}$ equals zero. Conversely, for R3, $H_c$ is determined exclusively from MR data, defining it as the field where MR exhibits its minimum (upturn). The data presented in b,c was obtained after the initial thermal and field cycling of the device. 
    }
    \label{fig:Fig5}
\end{figure*}

To demonstrate the magneto-electronic functionality facilitated by this approach, we conducted electronic transport assessments of the \device\ heterostructure that was presented in Figure 4b--e. Figure 5a shows the fabricated mesoscopic device (labeled D1), comprising three distinct measurement regions, including Fe-intercalated \HTaS\ (region R1), Fe-intercalated \HNbS\ (region R3), and their composite heterostructure (region R2) (Figure 5a). In regions R1--R3, the temperature-dependent longitudinal resistance reveals an inflection near 30 K (Extended data Figure 2), suggesting a phase transition upon cooling. Additionally, upon cooling in a 12 T magnetic field, these inflections broaden and shift to higher temperatures for R1--R2 and lower temperatures for R3 (Extended data Figure 2), implying the magnetic origin of the phase transition consistent with ferromagnetism\supercite{morosan2007sharp} in R1-R2 and antiferromagnetism in R3\supercite{Haley2020}. Further understanding of the long-range magnetic ordering in R1--R3 was gained through measurements of the magnetoresistance (MR) and field-dependent anomalous Hall resistance (\textit{R}\textsubscript{AHE}) (Figure 5b--c, Extended data Figure 3a--c), which reflect the interactions between the itinerant hole carriers and the magnetic moments. It is noteworthy that all regions exhibit a strong response to an out-of-plane magnetic field ($H \parallel c$) while showing negligible response to an in-plane field ($H \parallel ab$). This confirms their strong magnetocrystalline anisotropy, a characteristic imperative for stabilizing magnetism in the ultrathin limit. 

First, we turn to the detailed analysis of the magnetotransport of R1 and R3, which consist of a singular host lattice. Region R1 (Fe\textsubscript{0.32(1)}TaS\textsubscript{2}) manifests a hysteretic negative magnetoresistance with a bow-tie-shaped profile, characteristic of ferromagnets\supercite{morosan2007sharp,husremovic2022hard} (Figure 5b,c and Extended data Figure 3a). Moreover, the presence of hysteresis in its $R_{\mathrm{AHE}}$ signal, which is proportional to magnetization, further supports the ferromagnetic characteristics of R1\supercite{morosan2007sharp,husremovic2022hard} (Figure 5b,c and Extended data Figure 3a). Conversely, the magnetoresistance (MR) behavior of R3 (Fe\textsubscript{0.32(3)}NbS\textsubscript{2}) mirrors that of metallic antiferromagnets, demonstrating positive MR below the N\'eel temperature and transitioning to negative MR above 40 K\supercite{Haley2020,liu2022unconventional,yamada1973magnetoresistance} (Figure 5b,c and Extended data Figure 3c). However, contrary to the expected behavior for antiferromagnets, R3 also exhibits hysteresis in its MR profile below the ordering temperature, suggesting the presence of uncompensated magnetic moments. Given that our samples deviate from stoichiometric Fe\textsubscript{0.33}NbS\textsubscript{2}, we propose that these uncompensated moments arise from intercalant disorder, which is known to introduce an uncompensated spin glass phase in bulk crystals of Fe\textsubscript{$x$}NbS\textsubscript{2}\supercite{ic01-maniv2021exchange,maniv2021antiferromagnetic}. The spin glass phase could become pinned by the surrounding antiferromagnetic moments, leading to the observed hysteretic behavior.

Having established R1 as ferromagnetic and R3 as antiferromagnetic with a minority uncompensated phase, the transport behavior of the heterostructure (R2) was examined (Figure 5b,c and Extended data Figure 3b). The magnetotransport characteristics of R2 are dominated by the Fe\textsubscript{0.31(2)}TaS\textsubscript{2} segment of the heterostructure, aligning with the ferromagnetic behavior observed in R1 (Figure 5b,c and Extended data Figure 3a,b). However, following the initial thermal and field cycling, the coercive fields ($H_c$) of R2 consistently surpass those of R1 below the N\'eel transition of R3 (Figure 5c). This suggests that out-of-plane interactions across the ferromagnetic-antiferromagnetic heterostructure may be responsible for the high coercivity of R2. Consequently, the heterostructure behaves akin to an exchange spring magnet, a composite of dissimilar magnets whose interfacial interactions increase the coercivity\supercite{kneller1991exchange}. These results underscore the potential of intercalated heterostructures, which are enabled by combining nanoscale topochemical reactions and vdW heterostructure assembly, as a versatile platform for designing and interrogating low-dimensional magnetic phenomena.

\section*{Discussion}
This study demonstrates the successful synthesis of TMD heterostructures intercalated with open-shell transition metal ions by implementing vdW assembly and nanoscale solid-state topochemistry of the TMD with a transition metal oxide film. The intercalation process is comprehensively investigated for \FexTaS, using both bulk and atomic-scale structural and spectroscopic probes. We uncover that intercalation is thermally activated and accompanied by the disproportionation of oxide-exposed \HTaS\ layers. We observe that the intercalants diffuse from the region in contact with the oxide film, allowing for the fabrication of pristine few-layer magnetic intercalation compounds under hBN-encapsulated flakes. Leveraging these insights, we use patterning of metal carbonyl-derived oxide films to direct metal ions into well-defined vdW heterostructures, expanding the range of accessible low-dimensional magnetic intercalation compounds as well as producing magnetic multilayer systems with exceptionally clean surfaces and interfaces. Representative measurements of such heterostructures show transport behaviors signifying strong interfacial magnetic exchange effects that are only possible with atomically clean interfaces.

We envision the emergent physics of these heterostructures may be controlled by altering the thickness of constituent flakes and the intercalants' identity, stoichiometry, and homogeneity. Additionally, we expect that the synthesis of heterostructures with atomically sharp moir\'e heterointerfaces will offer exciting opportunities to both control the intercalation chemistry and also to fundamentally engineer complex magnetic behaviors. This customizability of vdW heteroepitaxy combined with nanoscale intercalation chemistry unlocks versatile opportunities to fine-tune the spin degree of freedom in materials, positioning intercalated heterostructures as compelling platforms for multi-component spintronic device architectures.

\newpage
\captionsetup[figure]{justification=justified,singlelinecheck=false,name={Extended Data Figure}}
\setcounter{figure}{0}

\begin{figure*}[htbp]
    \centerline{\includegraphics[width=3.5in]{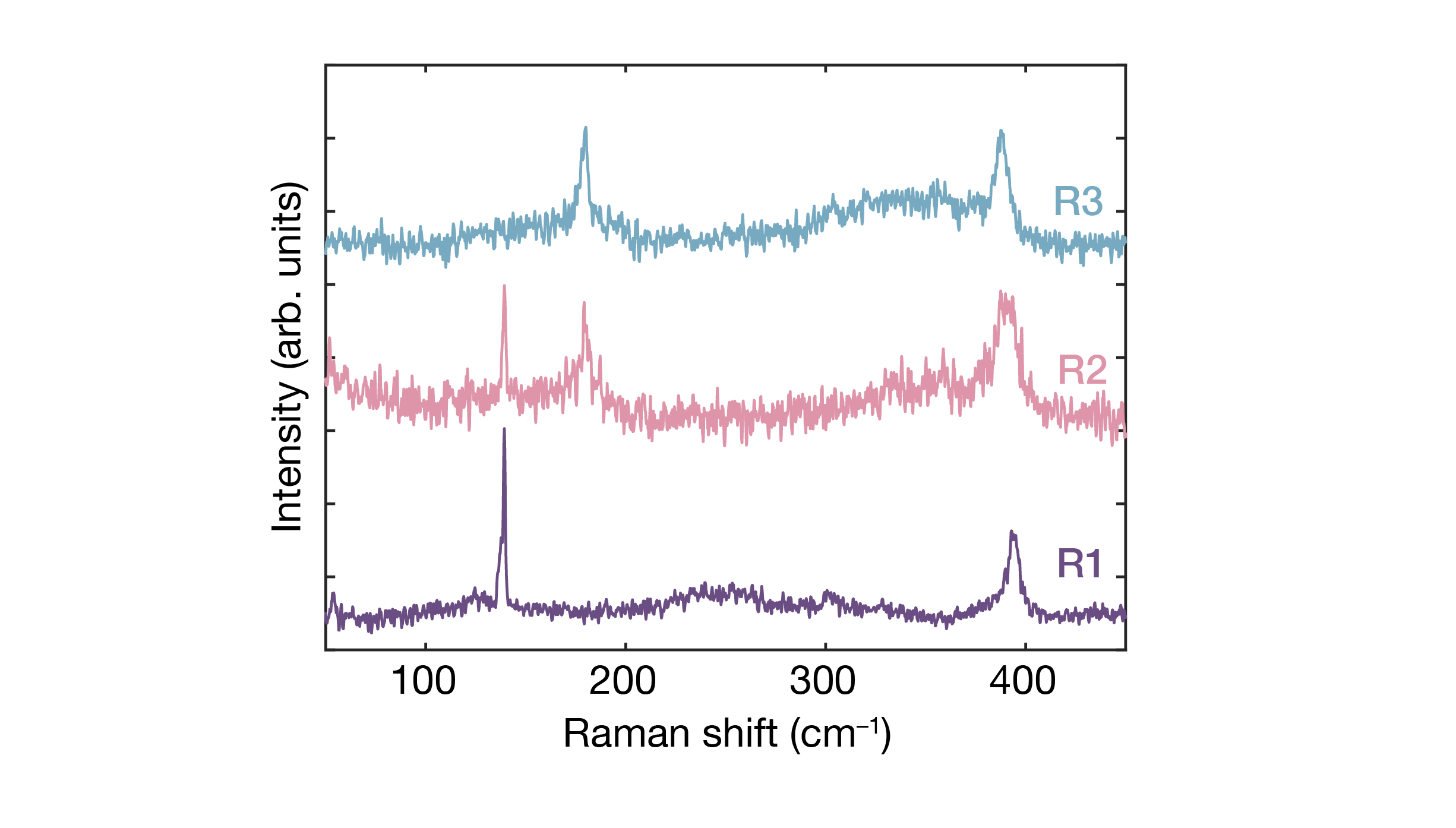}}
    \caption{\textbf{Raman of device D1.} Room-temperature Raman spectra of regions R1--R3 of device D1. 
    }
    \label{fig:Extended1}
\end{figure*}

\begin{figure*}[htbp]
    \centerline{\includegraphics[width=3.5in]{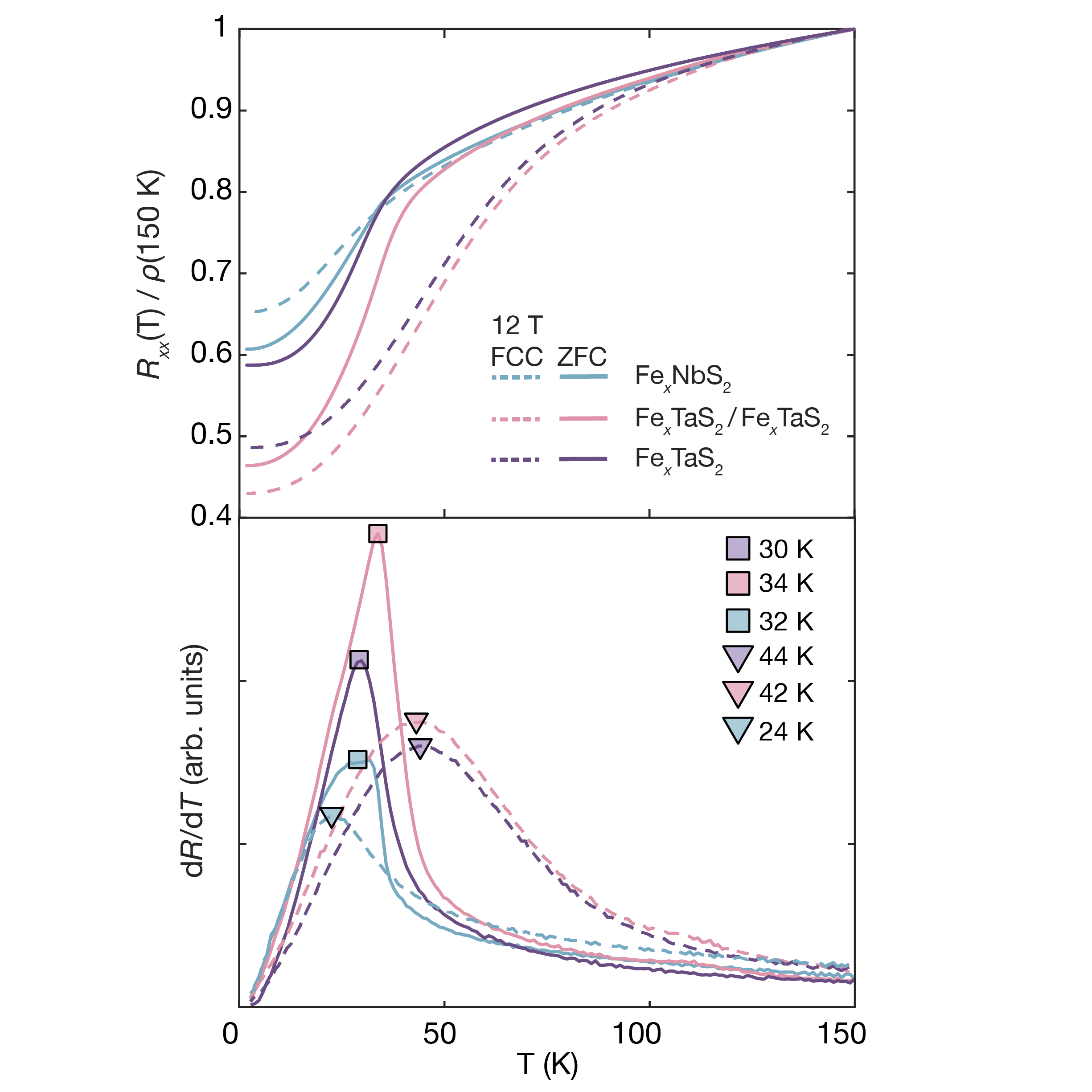}}
    \caption{\textbf{Temperature-dependent resistance of device D1.} Temperature-dependent resistance and its first derivative ($dR/dT$) for R1--R3 upon zero-field cooling (ZFC) and cooling in a 12 T magnetic field ($H \parallel c$). Maxima of $dR/dT$ curves are labeled. 
    }
    \label{fig:Extended2}
\end{figure*}

\begin{figure*}[htbp]
    \centerline{\includegraphics[width=\textwidth]{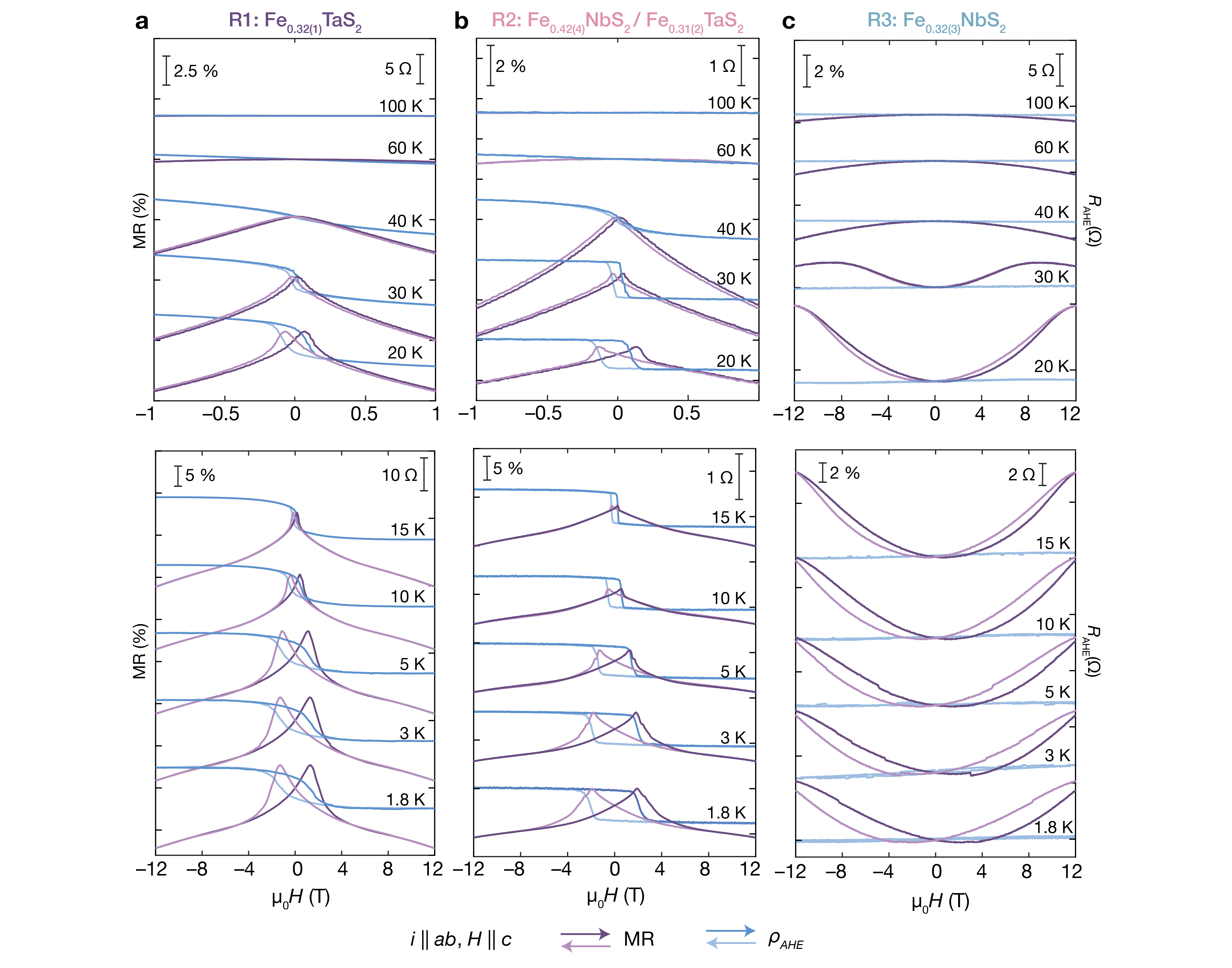}}
    \caption{\textbf{Temperature-dependent magnetotransport of device D1.} \textbf{(a--c)} Temperature-dependent magnetoresistance (purple) and anomalous Hall resistance (blue) for R1 \textbf{(a)}, R2 \textbf{(b)} and R3 \textbf{(c)}. The stoichiometries of iron relative to the Ta/Nb stoichiometry were determined through STEM-EDS analysis of cross-sections prepared subsequent to magnetotransport experiments. Data was obtained after the initial thermal and field cycling of the device.
    }
    \label{fig:Extended3}
\end{figure*}

\clearpage

\section*{Acknowledgements}
We thank R. Murphy and M.P. Erodici for helpful discussions, and we acknowledge C. Gammer and S. Zeltmann for developing the 4D-STEM acquisition code for TitanX. We also acknowledge S.S. Fender and G.P. Hegel for their help with bulk characterization. This work was supported by  the Gordon and Betty Moore Foundation EPiQS Initiative Award no. 10637. This research used resources of the Advanced Light Source, which is a DOE Office of Science User Facility under contract no. DE-AC02-05CH11231. Work at the Molecular Foundry, LBNL, was supported by the Office of Science, Office of Basic Energy Sciences, the U.S. Department of Energy under Contract no. DE-AC02-05CH11231.  Confocal Raman spectroscopy was supported by a Defense University Research Instrumentation Program grant through the Office of Naval Research under award no. N00014-20-1-2599 (D.K.B.). Electron microscopy was, in part, supported by the Platform for the Accelerated Realization, Analysis, and Discovery of Interface Materials (PARADIM) under NSF Cooperative Agreement no. DMR-2039380. This work made use of the Cornell Center for Materials Research (CCMR) Shared Facilities, which are supported through the NSF MRSEC Program (no. DMR- 1719875). The Thermo Fisher Spectra 300 X-CFEG was acquired with support from PARADIM, an NSF MIP (DMR-2039380) and Cornell University. K.W. and T.T. acknowledge support from JSPS KAKENHI (Grant Numbers 19H05790, 20H00354 and 21H05233). B.H.G. was supported by the University of California Presidential Postdoctoral Fellowship (PPFP) and the Schmidt Science Fellows, in partnership with the Rhodes Trust. L.S.X. was supported by an Arnold O. Beckman Postdoctoral Fellowship. S.H. acknowledges support from the Blavatnik Innovation Fellowship. 

\section*{Author Contributions}
S.H. and D.K.B. conceived the study. S.H., O.G., and W.Z. fabricated the samples. S.H., B.H.G., K.C.B, and C.S. imaged samples using TEM. S.H., L.X., O.G., S.H.R., C.J., A.B., E.R. performed nanoARPES experiments. Z.K. performed the XPS study. S.H., S.M.R., and C.O. developed the code for the virtual apertures. T.T. and K.W. provided the hBN crystals. S.H. and D.K.B. wrote the manuscript with input from all co-authors.

\section*{Competing Interests}
The authors declare no competing interests.

\section*{Additional Information}
Correspondence and requests for materials should be emailed to DKB\\
(email: bediako@berkeley.edu).

\printbibliography
\end{document}